\documentclass[cameraready]{Interspeech}

\title{AugCodec: A Low-Bitrate Disentangled Neural Speech Codec via Data Augmentation}

\author{Dongmei}{Wang}
\author{Xiaohang}{Sun}
\author{Yang}{Liu}
\author{Fanjie}{Kong}
\author{Abhishek}{Yanamandra}
\author{Abhinav}{Jain}
\author{Daniel}{Tompkins}
\author{Woohyun}{Kang}
\author{Najmeh}{Sadoughi}
\author{Sunil}{Hadap}
\author{Xiang}{Hao}
\author{Zhu}{Liu}
\author{Caren}{Chen}

\address{
    Amazon, USA
}

\email{dmwang@amazon.com}


\usepackage{comment}
\usepackage{booktabs}
\usepackage{multirow}
\usepackage{arydshln}
\usepackage{graphicx}

\begin{document}

\maketitle

\begin{abstract}
We propose AugCodec, a low-bitrate disentangled neural speech codec that leverages data augmentation to decompose speech into three distinct components: semantic, speaker, and prosody tokens. Specifically, we employ tailored augmentation strategies to transform speech into distinct variants, each serving as input for extracting tokens that preserve the target attribute while suppressing others. This disentanglement strategy enables substantial reduction in token rate. Furthermore, we introduce an augmentation loss that aligns semantic encoder outputs between source and voice-converted speech, encouraging speaker-agnostic embeddings while mitigating the acoustic mismatch induced by voice conversion. Experiments on LibriSpeech test-clean demonstrate that AugCodec significantly outperforms state-of-the-art methods in both reconstruction quality and disentanglement, while operating at only $12.5Hz$ with three token streams.
\end{abstract}

\section{Introduction}
Beyond compression for transmission, neural speech codecs have been extensively explored in recent years as foundational building blocks for speech language models and multimodal models \cite{soundstream, encodec, dac, li_codec_investi_2024, pooneh2025AudioTokensSurvey}. The research scope has expanded from an initial focus on reconstruction quality \cite{soundstream, encodec, dac} to encompass a broader range of aspects, including efficient token rates, disentanglement, streaming, lightweight design, and compatibility with speech language models at various levels \cite{xin2024bigcodec, semantiCodec, moshi_2024, Qwen3-TTS-2026,coreteam2025mimoaudio}.

Disentangled speech codecs encode speech into distinct types of tokens based on speaker attributes. FACodec \cite{NaturalSpeech3_2024} is one of the works that disentangles speech signals into prosody tokens, content tokens, acoustic detail tokens, and speaker embeddings. Although various augmentation losses are applied to facilitate disentanglement, all features are extracted from the same speech source, which limits the disentanglement capability. A similar problem occurs in other works \cite{discoSpeech_Li_2025, wang2025sparktts}.

Recently, speech codecs operating at low-bitrate and low frame rate have also been explored. FreeCodec \cite{freeCodec} proposes a codec system operating at 57 tokens per second that disentangles speech into content, speaker, and prosody tokens. However, the authors did not report the word error rate (WER) for reconstructed speech, and the voice conversion task exhibits a relatively high WER, indicating that the semantic representation may be insufficient under the current settings. A text-prompt-based single-stream codec model with a diffusion-based decoder is proposed in \cite{TaDiCodec_wang_2025}. Despite achieving competitive results with low bit rate, it relies heavily on the assumption of text input as a condition. Some recent works also operate at low frame rates, but they rely on residual vector quantization (RVQ), which requires at least 8 tokens per frame for representation \cite{moshi_2024, Qwen3-TTS-2026, FlexiCodec_Li_2026}. Additionally, these methods do not support disentanglement.
 
In this work, we propose a low-bitrate disentangled speech codec that thoroughly disentangles speech features, enabling more compact token representations. To achieve effective disentanglement, we introduce a data augmentation strategy that extracts semantic, prosody, and speaker tokens from distinct sources to minimize cross-feature interference. Specifically, semantic features are extracted from voice-converted speech to remove speaker information; prosody features are derived from low-frequency STFT components, which contain minimal speaker or semantic information; and speaker features are extracted from a different utterance of the same speaker. These features are independently quantized into three distinct token sets, then combined and fed into the decoder for waveform reconstruction. Furthermore, we introduce an augmentation loss to align semantic encoder outputs between source and voice-converted speech, mitigating conversion-induced mismatch. Comprehensive evaluations on LibriSpeech test-clean demonstrate the effectiveness of our proposed method.

















\begin{figure*}[ht]
  \centering
  \includegraphics[width=\linewidth]{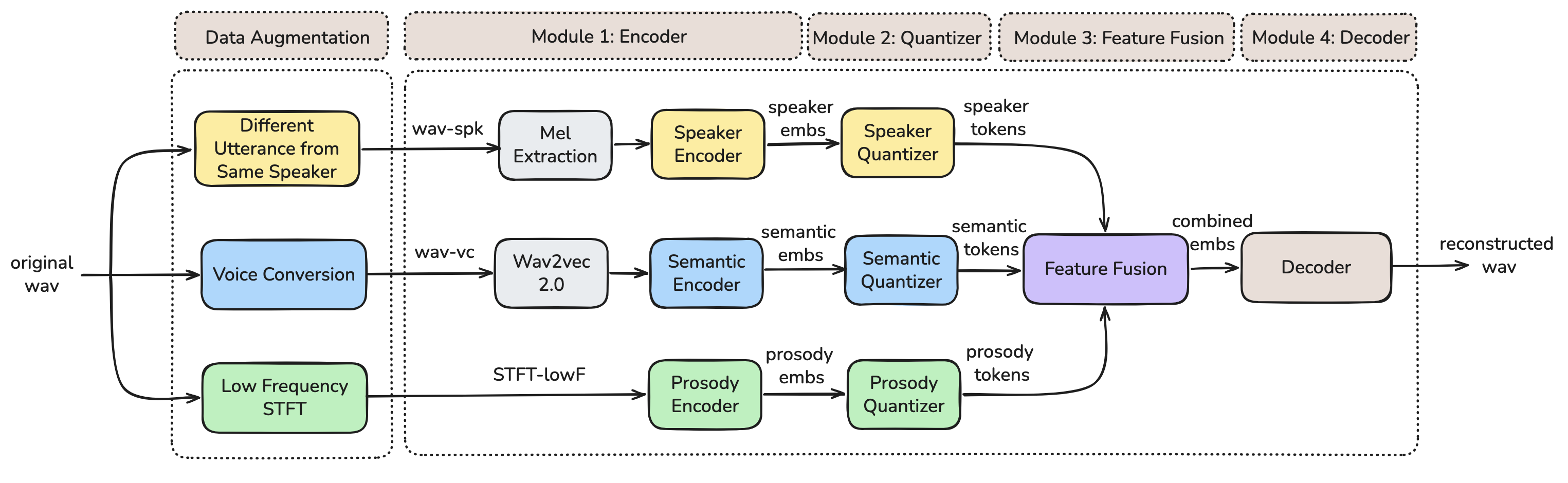}
  \caption{System overview of AugCodec during training.}
  \label{fig:model_architecture_AugCodec}
\end{figure*}

\section{AugCodec}
We introduce AugCodec, a low-bitrate disentangled neural speech codec. By leveraging data augmentation strategies, AugCodec decomposes speech signals into three distinct token types: semantic tokens, speaker tokens, and prosody tokens. This thorough disentanglement enables each feature dimension to be represented with fewer tokens, facilitating the development of a more compact and efficient low-bitrate speech codec.

\subsection{Data augmentation}
A key contribution of our work is the design of three specialized input variants that facilitate the disentanglement of semantic, speaker, and prosodic features during training, as shown in left side of Fig.~\ref{fig:model_architecture_AugCodec}.
For the semantic branch, a diffusion-based voice conversion model \cite{seedvc_2024} transforms the source speech to a different speaker, removing speaker characteristics while preserving only speaker-independent semantic information.
For the speaker branch, a different utterance from the same speaker is selected as input, thereby retaining speaker-specific characteristics while discarding content-related features.
For the prosody branch, only the lower-frequency components of the short-time Fourier transform (STFT) are retained from the original signal, thereby preserving the fundamental frequency (F0) and lower harmonics while removing the second or higher formant peaks.
Although some residual overlap between the semantic and prosody inputs may still exist, our approach is designed to minimize their correlation.
It should be noted that during inference, all features are extracted directly from the original source speech when performing reconstruction.

\subsection{Model architecture}
Right side of Fig.~\ref{fig:model_architecture_AugCodec} illustrates the overall model architecture of AugCodec, comprising four major modules: encoder, quantizer, feature fusion, and decoder. Prior to feature fusion, each feature dimension is processed independently through separate streams. 

\subsubsection{Encoder}
We employ three separate encoder streams to extract semantic, speaker, and prosody features. The semantic encoder adopts an architecture similar to BiCodec~\cite{wang2025sparktts}, with a key modification that introduces a compression scheme for further temporal reduction while preserving inter-frame dynamics. The speaker encoder follows the same architecture as BiCodec. The prosody encoder shares a similar architecture with the speaker encoder, augmented with a compression scheme.

\noindent \textbf{Semantic Encoder.} After obtaining the voice-converted (VC) speech from the original signal, we extract features using wav2vec 2.0, averaging the embeddings from the 11th, 14th, and 16th layers to serve as input. The input is first processed through ConvNeXt blocks~\cite{liu_convNeXt_2022} to jointly capture correlations along both the temporal and feature dimensions. The output of the ConvNeXt blocks retains the same temporal resolution as the input. Further compression is achieved by stacking consecutive frames along the feature dimension, followed by a linear projection to restore the original dimensionality:
\begin{equation}
\hat{\mathbf{z}}^{(t')} = \mathbf{W}{cpr} \left[ \mathbf{z}^{(t' r)} \parallel \cdots \parallel \mathbf{z}^{(t' r + r - 1)} \right] + \mathbf{b}{cpr}
\label{eq:compression-semantic}
\end{equation}
where $\mathbf{z}^{(t)} \in \mathbb{R}^{D}$ and $\hat{\mathbf{z}}^{(t')} \in \mathbb{R}^{D}$ denote the semantic embeddings before and after compression at time steps $t$ and $t'$, respectively; $t' = \lfloor \frac{t}{r} \rfloor$  is the compressed time index; $\parallel$ denotes the operation of concatenation; $\mathbf{W}{cpr} \in \mathbb{R}^{D \times (D \cdot r)}$ and $\mathbf{b}{cpr} \in \mathbb{R}^{D}$ are learnable parameters; and $r$ denotes the compression factor. Unlike average pooling, which inherently discards fine-grained distinctions across frames, this learned compression mechanism preserves temporal dynamics within the semantic content.

\noindent \textbf{Speaker Encoder.} For the speaker encoder, we follow~\cite{wang2025sparktts} and employ an ECAPA-TDNN architecture~\cite{desplanques_interspeech_2020} to extract frame-level speaker embeddings from the mel-spectrogram of a different utterance by the same speaker. These embeddings are subsequently aggregated into a fixed-length global representation through a cross-attention mechanism employing learnable queries.

\noindent \textbf{Prosody Encoder.} For the prosody encoder, we also employ ECAPA-TDNN architecture, but with different stride and kernel sizes to extract prosody embeddings. Since prosody information typically spans longer temporal segments, we use a coarser temporal resolution with a frame hop length of 160 ms.

\subsubsection{Quantization}

Quantization is performed on each feature stream independently. For the semantic stream, following~\cite{dac, wang2025sparktts}, we apply vector quantization \cite{vq-vae} to a lower-dimensional projection of the compressed semantic encoder embeddings to mitigate codebook collapse. For the speaker stream, we apply Finite Scalar Quantization (FSQ) \cite{FSQ_2024} on the global speaker embeddings to obtain the global speaker tokens. For prosody stream, we adopt the same FSQ quantization mechanism as speaker tokenization. 

\subsubsection{Feature fusion}
After quantization, the number of tokens from each feature stream may vary due to different encoding strategies and compression rates. These tokens are aligned and merged prior to decoding. 

\noindent \textbf{Semantic embedding expansion.} To reverse the temporal compression described in Equation~\eqref{eq:compression-semantic}, we apply an expansion operation that splits each compressed embedding along the feature dimension based on the expansion factor, and maps the segments back to the upsampled temporal positions via a linear projection:
\begin{equation}
\tilde{\mathbf{z}}^{(t \cdot r + i)} = \mathbf{W}_{exp} \hat{\mathbf{z}}^{(t)}_{i} + \mathbf{b}_{exp}
\label{eq:expansion}
\end{equation}
where $\hat{\mathbf{z}}^{(t)}_{i} \in \mathbb{R}^{D/r}$ denotes the $i$-th segment of the compressed embedding $\hat{\mathbf{z}}^{(t)}$; $\tilde{\mathbf{z}}^{(t \cdot r + i)} \in \mathbb{R}^{D}$ is the expanded embedding at time step $t \cdot r + i$; $i = 0, 1, \ldots, r-1$ is the segment index; $r$ is the expansion factor; $\mathbf{W}_{exp} \in \mathbb{R}^{D \times (D/r)}$ and $\mathbf{b}_{exp} \in \mathbb{R}^{D}$ are learnable parameters. Unlike naive interpolation-based upsampling, this learned expansion mechanism enables the recovery of fine-grained temporal variations.

\noindent \textbf{Speaker embedding expansion.} The quantized speaker embeddings are concatenated, projected to match the semantic embedding dimensionality, and repeated along the time dimension to match the source speech length.

\noindent \textbf{Prosody embedding expansion.} The quantized prosody embeddings are expanded in a similar manner to the semantic stream, but with a specific expansion factor to upsample from 160ms/frame to match the temporal resolution of the semantic embeddings. Additionally, a single transformer layer with positional encoding is applied to the expanded prosody embeddings to capture temporal dependencies.

\noindent \textbf{Feature merging.} The three disentangled embeddings are integrated through a multi-stage process. First, the expanded semantic embedding $\tilde{\mathbf{z}}^{(t)}$ and prosody embedding $\tilde{\mathbf{p}}^{(t)}$ are combined via element-wise multiplication:
\begin{equation}
\mathbf{h}^{(t)} = \tilde{\mathbf{z}}^{(t)} \odot \tilde{\mathbf{p}}^{(t)}
\label{eq:element-wise}
\end{equation}
where $\odot$ denotes element-wise multiplication. The resulting representation $\mathbf{h}^{(t)}$ is then modulated by the global speaker embedding $\tilde{\mathbf{s}}$ using FiLM-based adaptive layer normalization~\cite{peebles_scalableDiffusion_cvpr_2023, perez2018film} with a residual connection:
\begin{equation}
\mathbf{y}^{(t)} = \mathbf{h}^{(t)} + \left(\gamma(\tilde{\mathbf{s}}) \odot \text{LayerNorm}(\mathbf{h}^{(t)}) + \beta(\tilde{\mathbf{s}})\right)
\label{eq:film}
\end{equation}
where $\gamma(\cdot)$ and $\beta(\cdot)$ are learned linear transformations that produce the scale and shift parameters, respectively. Finally, a single-layer Transformer with positional encoding is applied to the fused features $\mathbf{y}^{(t)}$ to model temporal dependencies across time steps.

\subsubsection{Decoder}
\label{subsubsec:decoder}
The fused embeddings are fed into a decoder module to generate the output waveform. The decoder first applies a series of ConvNeXt blocks~\cite{liu_convNeXt_2022} to the fused embeddings. The remainder of the decoder follows an architecture similar to DAC~\cite{dac}, employing Snake activations~\cite{liu_snakeActi_NeurIPS_2020} for improved periodic signal modeling. It consists of an initial weight-normalized 1D convolution (kernel size 7), followed by a series of upsampling blocks. Each upsampling block comprises a Snake activation, a weight-normalized transposed 1D convolution for learnable upsampling, and three residual units with dilation rates of 1, 3, and 9. The decoder concludes with a Snake activation, a final weight-normalized convolution, and a Tanh activation to produce the output waveform.

\subsection{Training objectives}
The model is trained end-to-end using a combination of four loss functions: reconstruction loss, adversarial loss, quantization loss, and augmentation loss. The reconstruction loss consists of L1 loss computed on multi-scale mel-spectrograms and L1 loss computed on multi-scale STFT spectrum. The adversarial loss is derived from a multi-period discriminator for waveform discrimination and a multi-band multi-scale STFT discriminator. The quantization loss comprises both the codebook loss and the commitment loss, which encourage efficient utilization of the discrete codebook entries.

\noindent \textbf{Augmentation Loss.} We propose an augmentation loss to encourage the semantic encoder to learn speaker-agnostic representations while mitigating the mismatch introduced by voice conversion. Specifically, in addition to applying the semantic encoder to the voice-converted speech, we also apply it to the original source speech and minimize the L1 distance between the two resulting embeddings. This drives the semantic encoder to discard speaker-related information and focus on speaker-invariant content in the embedding space.

These losses are combined as a weighted sum to form the final training objective, with weights of 15.0, 5.0, 2.5, 1.0, 3.0, and 1.0 for the mel L1, STFT L1, adversarial-feature, adversarial-wav, quantization, and augmentation losses, respectively.

\begin{table*}[]
\caption{Reconstruction results on LibriSpeech-test-clean short splits ($4s$-$10s$)}
\label{tab:res_recons}
\centering
\resizebox{\linewidth}{!}{
\begin{tabular}{cccccccc}
\toprule
Systems                  & Frame rate & Bit-rate (bps) & Codebook size & WER $\downarrow$   & PESQ $\uparrow$ & SIM $\uparrow$  & UTMOS $\uparrow$ \\ \hline
GT                      & -           & -              & -               & 3.10  & -      & -     &  3.21     \\
\hdashline[1pt/2pt]\hdashline[0pt/1pt]
BiCodec                 & 12.5, global       & 312.50        & 8192, 4096        &  60.15     & 1.40     & 0.88     & 2.66      \\
Mimi                    & 12.5       & 412.50        & 2048 $\times$ 3        & 7.19  & 1.81 & \textbf{0.92} & 2.25  \\
Qwen-TTS-Tokenizer-12Hz & 12.5       & 412.50        & 2048 $\times$ 3        & 13.09 & 1.36 & 0.78 & 1.17  \\
\hdashline[1pt/2pt]\hdashline[0pt/1pt]
AugCodec-1         & 12.5, global, 6.25   & 362.50        & 2048, 4096, 4096        & 5.71  & 1.92 & 0.90 & \textbf{3.11}     \\
AugCodec-2         & 12.5, global, 6.25   & 387.50        & 8192, 4096, 4096        & 5.66    &1.94      & 0.90     & 2.93      \\ 
AugCodec-3         & 12.5, global, 6.25   & 400.00        & 16384, 4096, 4096       & \textbf{5.12}    & \textbf{1.99}      & 0.90     & 3.04     \\ 
AugCodec-4         & 6.25, global, 6.25   & 231.25        & 8192, 4096, 4096        & 17.11  & 1.67 & 0.88 & 2.78     \\ 
\hdashline[1pt/2pt]\hdashline[0pt/1pt]
\multicolumn{8}{c}{\textit{Ablation Study}} \\ 
\hdashline[1pt/2pt]\hdashline[0pt/1pt]
AugCodec-2 w/o $\mathcal{L}_{\text{aug}}$     & 12.5, global, 6.25   & 387.50   & 8192, 4096, 4096        & 6.29    &1.89      & 0.90     & 2.92  \\ 
\bottomrule
\end{tabular}
}
\end{table*}

\section{Experimental results}
\subsection{Experimental setup}
\subsubsection{Datasets} 
For training, we use the LibriLight-medium and LibriTTS datasets, resampled to 16kHz. For LibriLight-medium, we segment the data using the provided VAD labels and retain only segments between 2s and 15s, yielding approximately 3000 hours of training data. For evaluation, we select samples between 4s and 10s from LibriSpeech test-clean, resulting in 1237 samples.

\subsubsection{Configuration details} 
\noindent \textbf{Data augmentation:} We use Seed-VC \cite{seedvc_2024}, an off-the-shelf voice conversion method\footnote{https://github.com/Plachtaa/seed-vc}, to construct semantic encoder training data, with the output resampled to 16kHz. For prosody extraction, we retain only the STFT spectrum below 500Hz, discarding higher frequencies. The FFT size is set to 2048, yielding a 64-dimensional prosody encoder input. For speaker extraction, we crop or repeat the selected utterance to a fixed duration of $6s$. 

\noindent \textbf{Model architecture:} The semantic encoder takes 1024-dimensional wav2vec2 features at 50 frames per second as input. The encoder consists of 12 ConvNeXt blocks, with an input dimension of 1024, a hidden dimension of 384, an intermediate dimension of 2048, and an output dimension of 1024. The downsampling ratio is set to 1.0 to preserve the original temporal resolution. In the compression sub-module, we set the compression factor to 4, and the linear projection layer maps a 4096-dimensional concatenation input to a 1024-dimensional output. The speaker encoder takes 128-dimensional mel spectrogram as input and employs an ECAPA-TDNN module with a latent dimension of 128 and an output dimension of 1024. The prosody encoder also adopts ECAPA-TDNN, with an input dimension of 64, an output dimension of 1024, and a downsampling rate of 8. For the feature fusion module, the transformer layers used for both prosody and merged features share the same configuration: an input dimension of 1024, 16 attention heads, a feedforward dimension of 128, and a dropout rate of 0.1. For decoder, the first part consists of 18 layers of ConvNeXt blocks, with an input dimension of 1024, a hidden dimension of 1024, the intermediate dimension of 8192. The second part includes the upsampling layer already detailed in sec \ref{subsubsec:decoder}. 

\noindent \textbf{Optimization:} We use the AdamW optimizer \cite{loshchilov2019decoupled} with a learning rate of 1e-4, $\beta_1 = 0.8$, and $\beta_2 = 0.99$. The batch size is set to 72, and the model is trained for 750k iterations. A random 3-second segment is sampled from each utterance for training.

\subsubsection{Evaluation metrics} 
To evaluate reconstruction quality, we employ PESQ \cite{pesq}, word error rate (WER), speaker similarity (SIM), and UTMOS \cite{saeki2022utmos}. Specifically, WER is computed from transcriptions generated by Whisper-v3-large \cite{radford2023robust}, while speaker similarity is measured using the WavLM-base-plus-sv model \cite{Chen2021WavLM}, which is fine-tuned for speaker verification.

\subsubsection{Compared Methods}
We compare our method with following baseline neural speech codec systems: BiCodec \cite{wang2025sparktts}, Mimi \cite{moshi_2024}, Qwen-TTS-Tokenizer-12Hz \cite{Qwen3-TTS-2026} and FACodec \cite{NaturalSpeech3_2024}. BiCodec is originally designed to have single stream semantic tokens at a frame rate of 50Hz plus global speaker tokens, we retrain the model to make it as 12.5Hz by adding our compression module to achieve the compression. We keep the encoder and decoder identical between AugCodec and the retrained BiCodec. For Mimi\footnote{https://github.com/kyutai-labs/moshi}, Qwen-TTS-Tokenizer-12Hz\footnote{https://github.com/QwenLM/Qwen3-TTS} and FACodec\footnote{https://github.com/lifeiteng/naturalspeech3\_facodec}, we use their open source inference code and model for inference. 

\subsection{Reconstruction Results}
The experimental results are presented in Table~\ref{tab:res_recons}. In addition to reconstruction metrics, we report the frame rate, bitrate, and codebook size for each system. For frame rate and codebook size, values are listed per token type in the order of semantic, speaker, and prosody. For Mimi and Qwen-TTS-Tokenizer-12Hz, where all token types share the same configuration, we denote it as \textit{codebook size $\times$ number of codebooks}, and use only the first 3 quantization layers for fair comparison. For BiCodec and AugCodec, since speaker tokens are global rather than frame-level, their frame rate is denoted as ``global'' and the bitrate is approximated using the semantic token frame rate. Ground truth (GT) results are included for reference.

Table~\ref{tab:res_recons} demonstrates that AugCodec consistently outperforms all baseline systems across most reconstruction metrics under comparable bitrate and frame rate conditions. Increasing the codebook size for semantic tokens yields further WER improvements. BiCodec, despite sharing a similar model architecture, lacks data augmentation and prosody tokens, leading to significantly degraded performance. Mimi achieves the highest speaker similarity. Qwen-TTS-Tokenizer-12Hz underperforms across all metrics, partly due to the absence of quantization dropout during training. Notably, even at a reduced semantic frame rate of 6.25Hz (AugCodec-4), our method outperforms BiCodec at 12.5Hz across all metrics, demonstrating strong scalability to ultra-low frame rates. These findings collectively validate the effectiveness of our proposed disentanglement strategies. Furthermore, the ablation study shows that removing the augmentation loss leads to degradation in WER and PESQ, validating its effectiveness.

\subsection{Disentanglement results}
To further evaluate speaker disentanglement, we conduct voice conversion experiments on the LibriSpeech-test-clean short split dataset. We extract the speaker token from a randomly selected utterance of a different speaker as the target while retaining the semantic and prosody tokens from the source utterance for decoding. We compare AugCodec against FACodec \cite{NaturalSpeech3_2024}, BiCodec-50Hz (official open-source model\footnote{https://github.com/SparkAudio/Spark-TTS}), and our retrained BiCodec-12.5Hz model. As shown in Table~\ref{tab:res_vc}, AugCodec achieves the lowest WER at both 50Hz and 12.5Hz, demonstrating superior content preservation. Notably, AugCodec-50Hz outperforms FACodec-50Hz in WER while using only $60\%$ of the bitrate. The improvement is more pronounced at 12.5Hz, where AugCodec reduces WER from 65.43 to 5.87 compared to BiCodec, indicating substantially better speaker-content disentanglement with only a marginal trade-off in speaker similarity.

\begin{table}[]
\caption{Voice conversion results}
\label{tab:res_vc}
\centering
\resizebox{\linewidth}{!}{
\begin{tabular}{ccccc}
\toprule
Frame Rate & Systems          & Bit-rate (bps) & SIM $\uparrow$   & WER $\downarrow$    \\ \hline
\multirow{3}{*}{50Hz} & BiCodec     & 1250.00        & 0.86 & 4.60  \\
& FACodec     & 2400.00      & \textbf{0.90} & 3.76 \\   
& AugCodec   & 1450.00         & 0.85 & \textbf{3.19}  \\ 
\hdashline[1pt/2pt]\hdashline[0pt/1pt]
\multirow{2}{*}{12.5Hz} & BiCodec & 312.50       & \textbf{0.86} & 65.43 \\
& AugCodec & 362.50       & 0.85 & \textbf{5.87}  \\ 
\hdashline[1pt/2pt]\hdashline[0pt/1pt]
6.25Hz & AugCodec & 231.25       & 0.85 & 17.03  \\ 
\bottomrule
\end{tabular}
}
\end{table}

\section{Conclusion}
In this paper, we presented AugCodec, a low-bitrate disentangled speech codec comprising semantic, speaker, and prosody tokens, operating at a maximum frame rate of 12.5Hz. We achieved thorough feature disentanglement through tailored data augmentation that transforms speech into forms containing only target features while minimizing others. Additionally, we introduced an augmentation loss to align semantic encoder outputs between original and voice-converted speech, mitigating conversion-induced mismatch. Experiments on LibriSpeech test-clean demonstrated that AugCodec significantly outperforms state-of-the-art methods in both reconstruction quality and speech disentanglement. In future work, we plan to integrate AugCodec into downstream tasks such as text-to-speech and speech-to-speech translation.

\newpage

\section{Generative AI Use Disclosure}
Generative AI was used solely for minor language polishing and editing of the manuscript. No AI tools were involved in the research design, experimentation, or data analysis.

\bibliographystyle{IEEEtran}
\bibliography{mybib}

\end{document}